\documentclass[pdflatex,sn-nature]{sn-jnl}% Style for submissions to Nature Portfolio journals

\usepackage{graphicx}%
\usepackage{multirow}%
\usepackage{amsmath,amssymb,amsfonts}%
\usepackage{amsthm}%
\usepackage{mathrsfs}%
\usepackage[title]{appendix}%
\usepackage{xcolor}%
\usepackage{textcomp}%
\usepackage{manyfoot}%
\usepackage{booktabs}%
\usepackage{algorithm}%
\usepackage{algorithmicx}%
\usepackage{algpseudocode}%
\usepackage{listings}%
\usepackage{setspace}
\usepackage{bm}
\usepackage[none]{hyphenat}
\usepackage{amsmath} 
\usepackage{braket}
\newcommand{\vb}[1]{\mathbf{#1}}
\newcommand{\p}[1]{\left( #1 \right)}
\newcommand{\br}[1]{\left[ #1 \right]}
\newcommand{\cbr}[1]{\left\{ #1 \right\}}
\newcommand{\abs}[1]{\left| #1 \right|}

\definecolor{my_green}{RGB}{10, 230, 10} 

\theoremstyle{thmstyleone}%
\theoremstyle{thmstyletwo}%
\theoremstyle{thmstylethree}%
\begin{document}

\title[Article Title]{Active Quantum Nematics: The First Quantization}
\author[1]{\fnm{Ghansham R.} \sur{Chandel}}\email{ghansham@umd.edu}
\author[1]{\fnm{Siddhartha} \sur{Das}}\email{sidd@umd.edu}

\affil[1]{\orgdiv{Department of Mechanical Engineering}, \orgname{University of Maryland}, \orgaddress{\city{College Park}, \state{MD}, \country{USA}}}

%%==================================%%
%% Abstract %%
%%==================================%%

\abstract{ Nematic symmetry entails conserved quantized quantities such as number of topological defects and vorticity cells. Correspondingly, countless quantum analogies have been found in Active Nematics. We formalize Active Nematics and Liquid Crystal theory into the framework of Quantum Mechanics by introducing a complex valued Nematic Wavefunction to the Beris-Edward equations, thus splitting spatiotemporally varying nematic systems into quantized states. We obtain the Planck’s energy-frequency relationship for active micro-swimmers such as peristaltic worms and bacterium as a consequence of local complex phase-symmetry of the governing equations, similar to the gauge formulation of Electromagnetism. For organisms operating on diffusive chemotaxis, we obtain predator-prey dynamics that evolve to maximize/minimize pheromones field gradient overlap. Furthermore, when quantizing beating hearts, similar to the orbitals of hydrogen atoms, the state-function allows us to characterize hearts not only through the rhythm, but also the spaciotemporal distribution of contractile activity of various harmonics among healthy and unhealthy hearts.}




\maketitle
\section{Continuum or Quantum:}\label{sec1}

\begin{figure}[h]
\centering
\includegraphics[width=0.64\textwidth]{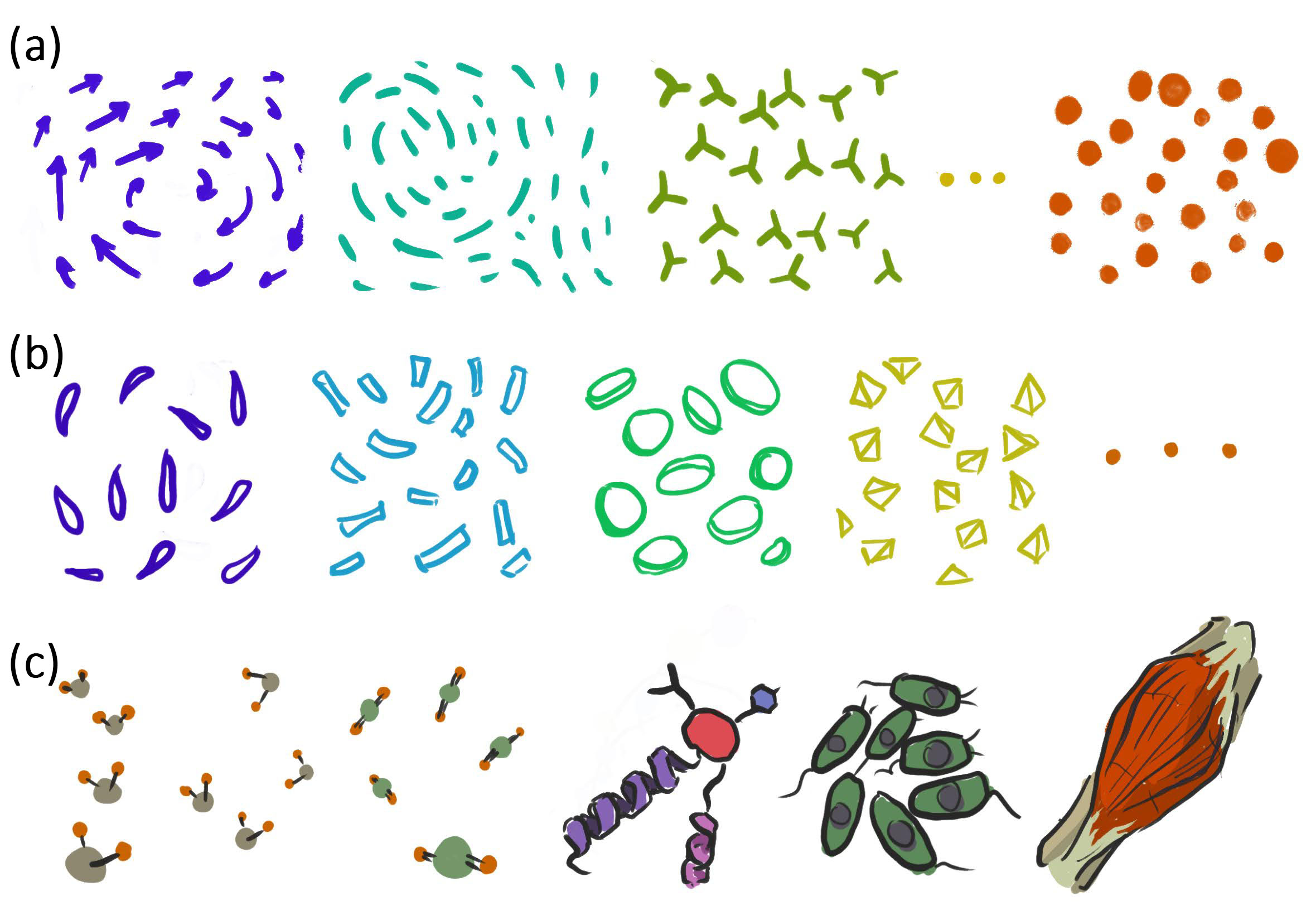}
\caption{ This figure illustrates the generality of continuums. (a) Shows mathematically idealized objects such as vectors, nematic director, tri-atic director, etc. (b) Shows idealized objects that may be represented by such mathematical objects. (c) Represents the objects in physical world that are routinely modeled as continumms.  
}
\label{fig1}
\end{figure}

We have nematic, tri-atic, four-atic, and in general, poly-atic particles \cite{PhysRevE.106.024701}, various polyhedrons, as well as rigid and soft chemical structures that may be packed densely to form continuums (illustrated in Fig. \ref{fig1}), which by the virtue of having non-zero temperature interact mutually by necessity. If the particles are sufficiently small in size and large in number, it is impossible to track individual particles and one must resort to the tools of statistical mechanics to describe averaged properties of the system. The transfer of these properties such as mass, momentum and energy across space and time give rise to continuum mechanics. Among such continuum systems, hydrodynamics is the generally adopted framework to describe thermo- \cite{cengel2022fundamentals}, electro- \cite{taylor1966studies}, magneto- \cite{alfven1942existence}, acoustic- \cite{laurell2014microscale}, as well as active matter phenomena \cite{marchetti2013hydrodynamics}. Interestingly, this framework in practice is distant with the other revolutionary continuum framework of quantum-mechanics, evidentially, due to a lack of complex numbers in applied hydrodynamics.

While conventionally the word continuum is in opposition to the term quantum, the complex field representing states of the system is continuous in space in many practical applications that follow from the Density Functional Theory. In classical quantum mechanics, the potential energy landscape builds stable modes of oscillations for the wavefunction and the total energy of the system puts the wavefunction in linear combination of those states in Schrodinger's framework. However, the two way coupling between the wavefunction and the potential energy landscape (coupling electromagnetic, strong, weak or gravitational forces) is natural in Dirac's relativistic formulation of quantum mechanics. Interestingly, in the theoretical and experimental literature pertaining to analogue blackholes, there is a precedence of treating acoustic waves in Navier-Stokes fluids as causal waves in Minkowski spacetime.

It would be a substantial leap in our understanding of physics if continuums made of symmetric/asymmetric objects could mimic the known laws of quantum mechanics at super-molecular scales. Hence, we constructively derive an interacting wavefunction and reformulate quantum mechanics in the reference frame of moving Galilean fluids (in section \ref{appendix}) and show how the fabric of the momentum field (or velocity $\vb{u}$) elegantly substitutes the Minkowski spacetime and even the electromagnetic fields while describing some of the most fundamental examples of interest in active matter physics. 

\newpage
\section{The Quantization of Active Nematics:}\label{sec2}
\subsection{Recap of Beris–Edwards Hydrodynamics}\label{subsec2_1}
Starting with the existing field-theoretic description, we glance at the Active Nemato-hydrodynamic equation, i.e., the Beris-Edward system of PDEs \cite{Yeomans08122025}\cite{marenduzzo2007hydrodynamics}. We define the velocity $\bf{u}$, the director field $\bf{n}$, the density $\rho$, and construct the simplest coordinate invariant object that possess a head-tail symmetry, i.e., a rank two, symmetric and traceless tensor with five independent fields $\vb{Q}(\vb{n})=q(\vb{nn}-\vb{I}/3)$ where $0\leq q\leq1$ is the scalar order parameter representing the degree of nematic ordering. The standard Active Nematics transport equations can thus be written as,

\begin{equation}
\partial_t \rho + \nabla\cdot\p{\rho\vb{u}}=0
,\label{BE1}
\end{equation}

\begin{equation}
\partial_t \p{\rho\vb{u}} + \nabla\cdot\p{\rho\vb{uu}}
=-\nabla p + \mu\nabla^2\vb{u}+\nabla\p{-\zeta\vb{Q}+\vb{\Pi}^e }
,\label{BE2}
\end{equation}

\begin{equation}
\partial_t \vb{Q} + \nabla\cdot\p{\vb{uQ}}-\vb{S}
=\vb{H}
,\label{BE3}
\end{equation}

\begin{equation}
\vb{S}\p{\partial_\vb{x}\vb{u},\vb{Q}}
=\lambda_\text{tb} \cbr{\vb{E},\vb{Q}+\frac{\vb{I}}{3}} +\br{\vb{Q}+\frac{\vb{I}}{3},\vb{\Omega}}
-2\lambda_\text{tb}\p{\vb{Q}+\frac{\vb{I}}{3}}\p{\vb{Q}:\nabla\vb{u}}
,\label{BE4}
\end{equation}

\begin{equation}
\vb{H}\p{\partial_\vb{Q}\mathcal{F}}=-\Gamma\frac{\delta\mathcal{F}}{\delta\vb{Q}}
,\label{BE5}
\end{equation}

\begin{equation}
\mathcal{F}\p{\vb{Q},\partial_\vb{x}\vb{Q}}= A_0\br{
\p{1-\frac{\lambda_0}{3}}Q_{ij}Q_{ij}
-\lambda_0 Q_{ij}Q_{jk}Q_{ki}+\lambda_0 \p{Q_{ij}Q_{ij}}^2
} +\p{\nabla Q_{ij}}^2\frac{L} {2}
,\label{BE6}
\end{equation}

\begin{equation}
\begin{split}
\Pi^e\p{\vb{Q},\nabla^2\vb{Q}} = 2\lambda_\text{tb} \p{\vb{Q}+\frac{\vb{I}}{3}}\p{\vb{Q}:\vb{H}}
+ \lambda_\text{tb}\br{{\vb{Q}+\frac{\vb{I}}{3}},\vb{H}}+\br{\vb{Q},\vb{H}} 
+ \nabla\vb{Q}\frac{\delta\mathcal{F}}{\delta\nabla\vb{Q}}
.\label{BE7}
\end{split}
\end{equation}

In eqs. (\ref{BE1}-\ref{BE7}), $\vb{S}$ is the alignment tensor given by the strain rate $\vb{E}$,  the vorticity tensor $\vb{\Omega}$ and a dimensionless tumbling parameter $\lambda_\text{tb}$ in a fluid of viscosity $\mu$ and density $\rho$. The gradients of the Landau-de Gennes (LdG) energy $\mathcal{F}$ gives the thermodynamic conjugate of $\vb{Q}$ as the molecular field tensor $\vb{H}$ governed by the nematic energy density $A_0$, a dimensionless $\lambda_0$ and the one-constant elasticity $L$. Furthermore, $\Gamma$ represents amount of nematic order stored per unit energy density that could be converted into mechanical work. Additionally, $\zeta$ is the activity constant and $\cbr{\cdot,\cdot}$ and $\br{\cdot,\cdot}$ are the standard anticommutator and commutator operators. Lastly, $\partial_\vb{x}$ and $\partial_{\vb{Q}}$ are the tensor valued gradients (covectors) represented compactly and are defined for each index of the varying mathematical object such as $x_i$ and $Q_{ij}$. Lastly, eq. (\ref{BE2}) is shown to have non-linear convective terms, not because those are relevant for the dynamics, but because those terms elucidate the underlying structure of material transport.

We denote the spatial coordinates with $\vb{x}$ and material coordinates with $\vb{X}$ and use $\partial_{Xt}$ to represent a generalized material derivative in Galilean, fluid filled systems as $\partial_{Xt}=\partial_t+\nabla\cdot\vb{u}\_$ for scalar and vectors and $\partial_{Xt}=\partial_t+\nabla\cdot\vb{u}\_+\hat{\vb{S}}|_\vb{u}\p{\_}$ for tensor fields. Here $\hat{\vb{S}}$ represents the operator that rotates any given tensor (such as $\vb{Q}$) as the background fluid ($\vb{u}$) shears and rotates. Notice that eqs. (\ref{BE1}-\ref{BE3}) are the governing equations, eqs. (\ref{BE4}-\ref{BE7}) are definitions and, eqs. (\ref{BE5}-\ref{BE7}) are model specific expressions, open to change from a model to another. The eq. (\ref{BE3}) is of the form $\partial_{Xt}\vb{Q}=-\delta\mathcal{F}/\delta\vb{Q}$ for some functional $\mathcal{F}$, i.e., $\vb{Q}$ follows gradient dynamics, whereas eq. (\ref{BE2}) has no such terms. Furthermore, both viscous ($\mu\nabla^2\vb{u}$) and passive elastic ($\vb{\Pi}^e$) stresses are restoring, i.e., passive liquid crystals are always pulled downwards in the energy landscape by increasing entropy. Hopes of any self-sustained, spontaneous dynamics would be lost if not for the active forces, which constantly pump mechanical energy into the system. 

\newpage
\subsection{The Governing Equations}\label{subsec2_2}

In this section, we focus on the implications and understanding of the resulting equations. The crux of the discretization lies in the abundance of wave-like patterns among innumerable animal kingdoms at molecular, cellular, tissue, organ, and herd levels of scales. Kinesin motors propelling microtubules, microbes swimming through flagellating cilia and flagella, colon transporting digestive materials, etc. are examples where nature settled on repeating, non-reciprocating motions, broadly label as peristaltic motion. Repeating (possibly peristaltic) motions can be mapped onto circles in the complex plane whose area scales with the amount of ``alive-matter", $\rho_\psi=\abs{\psi}^2$. With the local angular speed and the wave-vector associated with active patches as $\omega\p{\vb{x},t}$ and $\vb{k}\p{\vb{x},t}$, the evolution of the wavefunction $\psi$ is given by eq. (\ref{q1}) (see section \ref{appendix1}):

\begin{equation}
    \p{\partial_{t} + \nabla\cdot\vb{u}} \psi= \text{i}\p{\omega_0 + \omega + \vb{u}\cdot\vb{k}}\psi
    \label{q1}
\end{equation}

Here, $\vb{u}$ is the spatial velocity. Physically, $\omega$ assesses how quickly the biological ``clock'' ticks and $\vb{k}$ assesses how further apart the phase repeats when moving with $\vb{u}$ at any given time, thus governing the overall evolution of $\psi$. Reasonably, if a creature carries out its bodily function faster (larger the $\omega$ for the same amplitude of $\psi$), more energy the creature consumes (larger the $E$). Therefore, energy-rate of an eigenstate of constant $\omega$ is $E=\hbar_A\omega$ where $\hbar_A$ is a creature specific, active-matter Planck constant. Straightforwardly, the operator that has a constant $\omega$ as its eigenvalue is $\hat{E}=-\text{i}\hbar_A\partial_t$. A resting creature spends energy at basal metabolic rate ($\text{BMR}$) and at least complex creatures have a heartbeat ($\text{BPM}$). Thus, the Planck's relationship defines $\omega_0$ as $\hbar_A\omega_0=\text{BMR}\times\text{BPM}$.

Similarly, creatures with more moving parts (more fins/cilia or longer tails/flagella) must produce proportionally more swimming force (momentum-rate) for a constant frequency. As viscous drag dominates the inertial force at low Reynolds numbers (Re), a constant momentum-rate is equivalent to a constant speed of swimming, and a momentum-rate operator $\hat{\vb{P}}=-\text{i}\hbar_A\nabla$ is appropriate to define. Eq. (\ref{q1}) can thus be rewritten as: 

\begin{equation}
    \p{\hat{E}+\hat{\vb{P}}\cdot\vb{u}}\psi=\br{ E_0 + E +\vb{u\cdot\vb{P}}}\psi,
    \label{q1_2}
\end{equation}
where, $E_0=\hbar_A\omega_0$ In eq. (\ref{q1_2}), by observation, $\hat{\vb{P}}$ and $\vb{u}$ commute only if $\vb{u}$ is divergence free. For incompressible creatures swimming in incompressible fluids, this commutativity is straightforward. However, for describing the internal flow of spongy creatures, the distinction in the order of operation is necessary. Since any complex field can be represented as $\psi(\vb{x},t)=\abs{\psi}\text{e}^{{\text{i}\phi\p{\vb{x},t}}}$, we must avoid labeling $ \mathbb{R}\text{e}\p{\psi} $ as the physically meaningful part and embrace treating $\phi\p{\vb{x},t}$ and $ \abs{\psi}=\sqrt{\rho_\psi} $ as knobs that tune the fundamental fields. 

To quantize eq. (\ref{q1}), realize it is linear in $\psi$ and we are free to split $\psi$ into arbitrary chunks of space and time. The splitting undertakes immense meaning if it corresponds to eigenmodes of any physical operator. First, we establish the role of $\abs{\psi}$ in modeling active matter hydrodynamics. We modify the ($\rho,\vb{u},\vb{Q}$) evolution equations (eq. (\ref{BE1}-\ref{BE3})) for dynamics with heterogeneous distributions of active materials (a creature, cells or flocks represented by $\psi$) in fluid filled systems: 
\begin{equation}
    \partial_{Xt}\rho=\psi^*\hat{\dot{\rho}}\psi + f_{\phi \rho},
    \label{q2}
\end{equation}

\begin{equation}
    \partial_{Xt}\rho\vb{u}=-\nabla p+\mu\nabla^2\vb{u}+\mu_2\nabla\p{\nabla\cdot\vb{u}} +\rho_\psi c_D\vb{u}_A +\vb{f}_{\phi u} +\nabla\cdot\p{  \rho_\psi\vb{\Pi}^e-\psi^*\hat{\zeta}\psi\vb{Q} } , 
    \label{q3}
\end{equation}

\begin{equation}
    \partial_{Xt}\vb{Q}=\vb{H}+\vb{f}_{\phi Q}.
    \label{q4}
\end{equation}

Now, eq. (\ref{q2}) is eq. (\ref{BE1}) with an additional an source term $\dot{\rho}=\psi^*\hat{\dot{\rho}}\psi$. Furthermore, eq. (\ref{q3}) is eq. (\ref{BE2}) with modified viscous, active and passive stresses as $\mu_2\nabla\p{\nabla\cdot\vb{u}} + \rho_\psi c_D\vb{u}_A$, $\nabla\cdot\p{\psi^*\hat{\zeta}\psi\vb{Q}}$ and $\nabla\cdot\p{  \rho_\psi\vb{\Pi}^e}$ respectively. And $f_{\phi\rho}$, $\vb{f}_{\phi u}$ and $\vb{f}_{\phi Q}$ are the interaction terms described in section \ref{subsec2_3} If the amount of active matter in the system is spatiotemporally varying, $\vb{u}$ may diverge even at low Reynolds numbers and create drag through $\mu_2$ which relates to the second viscosity coefficient $\lambda$ as $\mu_2=\mu+\lambda$. Furthermore, closely packed particles are capable of dragging fluids and an additional Darcy drag term $\rho_\psi c_D\vb{u}_A$ with a drag coefficient $c_D$ of the active flocks moving with $\vb{u}_A$ relative to $\vb{u}$ is appropriate to add. One could model the drag coefficient to be a function of the orientation (nematic director $\vb{n}$) and distinguish even granular dynamics, but we assume it constant for simplicity. Guaranteed by the spectral theorem, the Hermitian operators corresponding to the growth rate $\dot{\rho}\p{\vb{x},t}$ or the activity $\zeta\p{\vb{x},t}$ can always be defined such that $\dot{\rho}=\psi^*\hat{\dot{\rho}}\psi$ and $\zeta=\psi^*\hat{\dot{\zeta}}\psi$ . Meaning, if $\psi$ is in superposition of multiple eigenstates, the basis vectors can be transformed to eigenstates of concerning quantities through respective operators. Example, for discrete or continuous eigenvalues of $\hat{\zeta}$ as $\zeta_n$ and $\zeta\p{\vb{x}_0,t_0}$  respectively,
\begin{align}
\begin{cases}
    \ket{\psi}= \underset{n}{\sum}c_n\ket{\psi_n} \\
    \ket{\psi}=\psi\p{\vb{x}_0,t_0}\ket{\delta\p{\vb{x}_0-\vb{x},t_0-t}} 
\end{cases} \Rightarrow 
\begin{cases}
\hat{\zeta}\ket{\psi} = \underset{n}{\sum} c_n \zeta_n \ket{\psi_n}\\
\hat{\zeta}\ket{\psi}=\zeta\p{\vb{x}_0,t_0}\ket{\psi}
\end{cases},
\label{eq_op}
\end{align}
where $\abs{c_n}^2$ and $\abs{\psi\p{\vb{x}_0,t_0}}^2$ describe the fraction of the system in the respective eigenstate. At zero energy and momentum rate, eq. (\ref{q1}) reduces to a material transport of $\psi$ and from eq. (\ref{eq_op}), the operators are expressed simply as $\hat{\dot{\rho}}=\dot{\rho}/\rho_\psi$ and $\hat{\zeta}=\zeta/\rho_\psi$. Hence, the quantum system is purely a fluid dynamical system. One might stretch to call this state an analogue Bose-Einstein of Active Matter where the entire state of the system is captured by a single lowest `energy-momentum-rate' state. This state is $E_0=\text{BMR}=\hbar_A\omega_0=m_0c^2$ where $m_0=\hbar_A/c^2$ is the BMR scaled to the units of mass-consumption-rate by the speed of sound in the fluid $c$.

\newpage
\subsection{Changing Reference Frames in AQN}\label{subsec2_3}

So far, we have highlighted the role of $\abs{\psi}$ in modeling spatial variations of relevant quantities in our governing equations, eqs. (\ref{q1}-\ref{q4}). The dynamics of the system on the other hand emerge from  spatiotemporal variations of the phase angle $\phi(\vb{x},t)$. Inactive patch floating in a background flow $\vb{u}_0$ have $\psi=\abs{\psi}\text{e}^{\phi_0}$, $\omega=0$ and $\vb{k}=\vb{0}$ for some constant $ \phi_0$.  When the patch becomes active, the creature reads the energy landscape and generates a compass $\phi(\vb{x},t)$ to move in the direction of maximum utility. Physically, this could mean following (avoiding) bio-signals for food (toxins). $\phi(\vb{x}_0,t)$ tells the temporal frequency of the motion and $\phi(\vb{x},t_0)$ dictates the wavelength of peristaltic motion which the body takes at $\p{\vb{x}_0,t_0}$ as seen in eq. (\ref{q_int}). Furthermore, similar to the Dirac equation, eq. (\ref{q1}) may be written as:

\begin{equation}
    \p{\,\text{i}\hbar\partial_\nu u^\nu+ u^\nu A_\nu + m_0c^2}\psi=0,
    \label{dirac}
\end{equation}
where $\partial_\nu=\cbr{\partial_t,c\nabla}$, $u^\nu=\cbr{1,\vb{u}/c}$ and $A_\nu=\hbar_A\cbr{\omega,c\vb{k}}$ are the augmented four-derivative, four-velocity and energy-momentum-rate vector-fields. In the spacetime picture, active creatures never move ($\vb{u}_A$) fast enough to change temporal velocities as the Lorentz factors $\gamma\p{\vb{u}_A}\approx 1$ and $\beta\p{\vb{u}_A}\approx\vb{u}_A/c$, i.e., Galilean transformations are sufficient. 

Activating creatures in AQN is same as giving fundamental particles an energy-momentum boost in quantum-field theory. Meaning, boosting the wavefunction $\psi=\abs{\psi}\text{e}^{\phi_0}\rightarrow\abs{\psi}\text{e}^{\phi_0+\phi\p{\vb{x},t}}$ in AQN is analogous to applying gauge transformation to the electromagnetic field after Lorentz boosting the spinor particles. Correspondingly, (see section \ref{appendix2} for derivation) the body (or its parts) take on non-zero energy-momentum rate $k_\nu=\cbr{\omega,c\vb{k}}$ and the interaction terms become non-zero:

\begin{equation}
\begin{cases}
    \omega=\partial_t\phi, \ \vb{k}=\nabla\phi,\\
    \vb{u}_A=\alpha\nabla\phi, \ \vb{u}=\vb{u}_0+\vb{u}_A,\\
    f_{\phi\rho} = \rho\alpha\nabla^2\phi, \\
    \mathbf{f}_{\phi u} + \rho_\psi c_D \vb{u}_A = \alpha\nabla[(\rho\partial_t - (\mu + \mu_2)\nabla^2)\phi]  + \rho_\psi q_\psi\nabla\phi,\\
    \mathbf{f}_{\phi Q} = \alpha\nabla \cdot (\mathbf{Q}\nabla\phi) + \hat{\mathbf{S}}|_{\alpha\nabla\phi}(\mathbf{Q}).
\end{cases}
\label{q_int}
\end{equation}

Here, $\alpha$ is the coupling strength between the creature's machinery and the fluid flow generation.  In eq. (\ref{q1_2}) we claimed and eq. (\ref{eq_op}) concludes that the creature's 'momentum' is directly proportional to the wavenumber ($\vb{u}_A=\alpha\vb{k}$), not only inside the active patch but throughout the universe. Interestingly,  $\rho_\psi q_\psi\nabla\phi$ is the term that pulls $\psi$ towards the direction of maximum change of $\phi$ with strength $q_\psi=\alpha c_D$ by creating $\vb{u}_A$. Under various assumptions and various regimes, one is free to drop various terms in eq. (\ref{q_int}), for example, far from the creature, the velocity can be assumed to be divergence free and terms involving $\nabla^2\phi$ may be ignored, or even dropping terms involving $\rho$ for low Re flows.

This completes our formulation of AQN. The magnitude and phase of the wavefunction is solved by coupled eqs. (\ref{q1_2}-\ref{q4}). Dividing motion into states of given frequency and given spatial distributions is basically a 3+1D Fourier transform of Active Matter. Lastly, boosting or activating creatures is same as changing the reference frame between stationary and moving creatures through appropriate gauge transformations in the fundamental fields. We are now ready to construct a few fundamental objects in this universe. 

\newpage
\section{Quantum Nematic Objects:}\label{sec3}
\subsection{Peristaltic Swimmers and Stationary Shakers}\label{subsec3_1}

At microscopic scale, between each propelled pair (or collection) of particles (shakers), the net momentum is always conserved as no external force acts on the bulk of active systems. This is in contrast with the “movers” such as ciliated bacteria, swimming sperms, worms and fishes that use internal energy to generate directional motion with no external force. Classically, this distinction is manifested in the ‘Scallop theorem’, i.e., swimmers using purely "reciprocal" motions cannot achieve net forward movement at vanishing Reynolds numbers. 

\begin{figure}[h]
\centering
\includegraphics[width=0.72\textwidth]{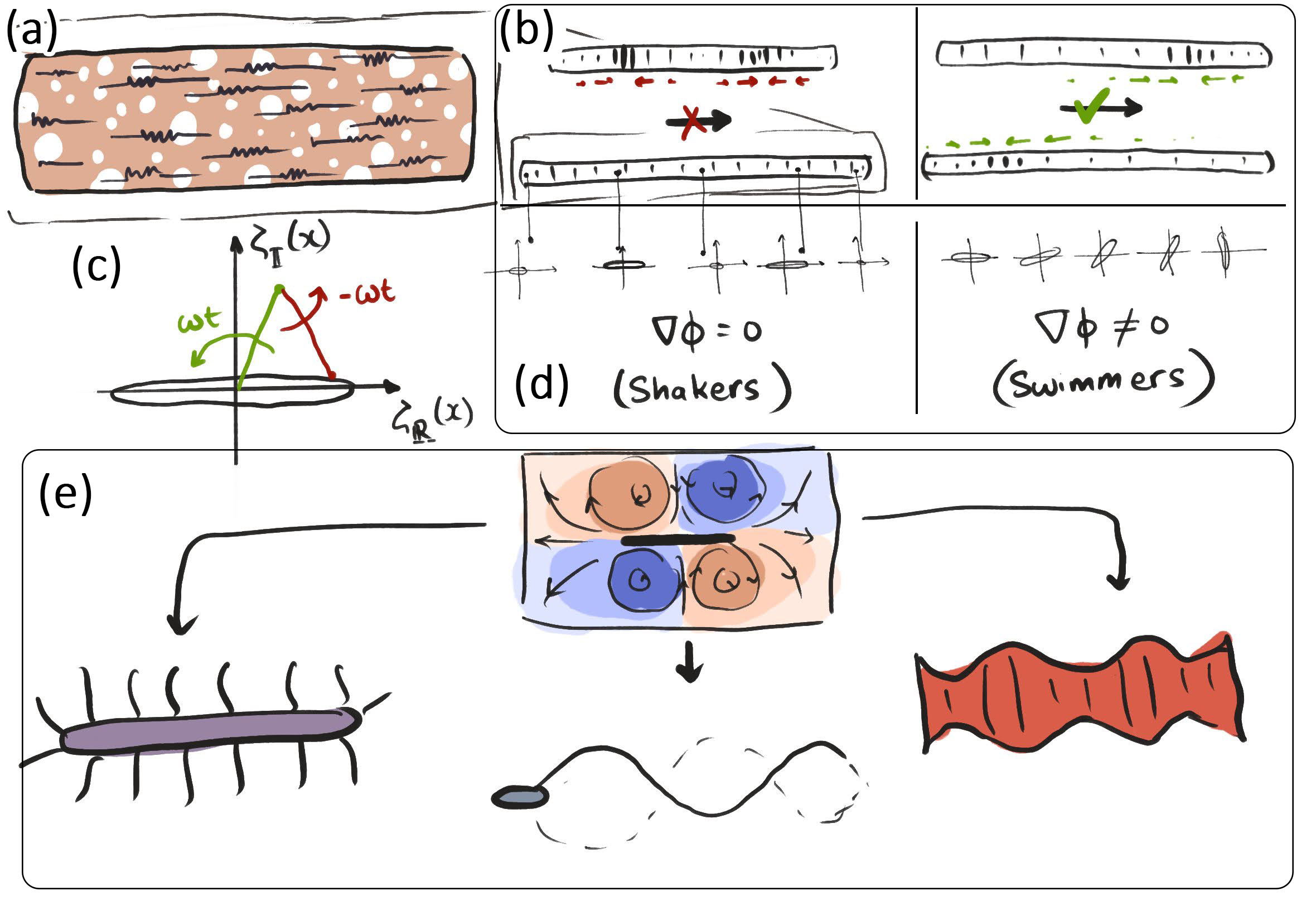}
\caption{ This figure illustrate the difference between peristaltic swimmers and shaker creatures. (a) Prototypical elongated active clump that has potential to swim. (b) When creatures moves its entire body in sync, no motion of self can be produced, but locally varying activation time induces peristaltic motion. (c) The motion of $\ket{\zeta}$ in the complex plane. (d) The visualization of $\phi$ with space that does not and dose induce motion. (e) Illustration of typical creatures capable of creating vortices for the purpose of movement.  
}\label{fig2}
\end{figure}

Assume a patch of region $V$ at origin generating active stress with strength $q_0\zeta(\mathbf{x}, t)$ in the $\mathbf{n} = \hat{\mathbf{x}}_1$ arrangement, where $q_0 = 1 + 3\sqrt{1 - 8/3\lambda_0}$ is the mean field scalar order \cite{marenduzzo2007steady}. For simplicity, we ignore the passive elastic stresses and assume no active particles are present outside the patch $V$, meaning, $\psi(\mathbf{x}, t) = 0$ for $\mathbf{x} \notin V$. Notice that eq. (11) is identically satisfied for $\mathbf{Q} = q_0(\hat{\mathbf{x}}_1\hat{\mathbf{x}}_1 - \mathbf{I}/3)$ as all spatial derivatives of $\mathbf{Q}$ are zero and $q_0$ is a root of polynomial terms and $\hat{\mathbf{x}}_1 \approx \hat{\mathbf{X}}_1$ for slender bodies. The basis vectors for this patch, which we call our creature are chosen to be of type $\psi_{n_xn_t} = \sin(\pi n_x x_1/l_0 + \theta_0)\mathrm{e}^{i n_t \omega t}$ as the standard solution of a particle in a box where $0 \le n_x < \infty$ and $n_t \in (-\infty, \infty)$ are integers which dictate the momentum and kinetic energy of this object respectively. Note that the value of $n_x/l_0$ is a system parameter as chosen by the creature in question. The $\theta_0$ is an arbitrary starting angle and length $l_0$. If the creature has eyes at the very front, it might be advisable to adjust a node at the tip through $\theta_0$ at $t = 0$ and the framework (eqs. (8-11)) readily transports the node (or any vestigial structures, such as eyes) perfectly through fluid advection.

It is sufficient to solve for $n_x = 1$ and $n_t = \pm 1$ as the functions are self-similar and can be stacked together. If our patch is a sponge with active muscles, a quick inspiration from peristaltic motion of worms helps us state that $\psi_{n_xn_t}$ are eigenstates of $\hat{\zeta}$ and $\hat{\dot{m}}$ with $\hat{\zeta}_{1,\pm 1} = \zeta$ and $\hat{\dot{m}}_{1,\pm 1} = -\rho\dot{\epsilon}$ (zero for any other quantum numbers), where $\dot{\epsilon}$ is the porosity rate of change due to sponge deforming under active stress. While an exact relationship between $\zeta$ and $\dot{\epsilon}$ may be found in terms material constants by solving the relevant porous elasto-hydrodynamics, we take them as prescribed quantities. Notice that the motion of this patch becomes a bouncing worm, because the active force distribution, given by $\mathbf{f}_\zeta = q_0\partial_{x_1}(\psi^*_{n_xn_t}\hat{\zeta}\psi_{n_xn_t})\hat{\mathbf{x}}_1$ equals (summation over repeated indices are implied):

\begin{equation}
    \mathbf{f}_\zeta = \frac{2\pi q_0\zeta}{l_0}\sin\p{\frac{2\pi \vb{x}}{l_0} + 2\theta_0}\cos{\omega t}.
    \label{a1}
\end{equation}

This integrates to non-zero value for $l_0\neq l_{\text{worm}}$ but average outs to zero for each $\Delta t=2\pi/\omega$, which shall come as no surprise as there is no spatial variation to the complex phase $\phi(\vb{x},t)=\phi(t)=n_{t}\omega t$. Furthermore, since eqs. (\ref{q1}-\ref{q3}) are linear in $\psi$ and $\vb{u}$, we can superimpose any number of $\psi_{n_xn_t}$ to create all reciprocating motions, but the active force totaled over each and any state $\vb{f}_\zeta=0$. This is the quantum analogue of the ‘Scallop theorem’. 

When we boost the wavefunction by a spatial gradient $\phi\p{\vb{x},t}=\phi(t)+k_{x_1}x_1$, an additional non-zero interaction force is generated on our body as $\vb{f}_{\phi u}=-q_\phi\rho_\psi\nabla\phi$ and $\vb{F}_{\phi u}=-q_\phi m_\psi\nabla\phi$, where $k_{x_1}$ is the wavenumber and $m_\psi=\rho_\psi V$ of the creature. Physically, the trajectory over time of $\ket{\zeta}$ follows a degenerated ellipse in the Hilbert space of the tensor product of all of our quantum states (two states corresponding to $n_t=\pm 1$) and rotates with angular speed $\omega$. Addition of $\phi(\vb{x})$ tells the creature to follow the same trajectory, but start the clock at different times over its length—inadvertently ending up imparting a finite mean to the otherwise purely oscillatory force. Remember, in the Stokesian limit, bodies under constant forces achieve terminal velocities, i.e., $\braket{\vb{u}}\propto\nabla\phi$. The momentum of the creature swimming in a viscous media becomes $\braket{\vb{p}}=\braket{\rho\vb{u}}$ and the de Broglie equation $\braket{\vb{p}}\propto k_{x_1}$ follows.

\newpage
\subsection{Predator and Prey Swimmers}\label{subsec3_2}

While deriving the interaction terms $f_{u\phi}$ and $\mathbf{f}_{\phi u}$, we highlighted that any $\phi_u$ is an external and $\phi$ is the internal potential driving the flow. The distinction however, is a matter of convention. With source potential labeled as $\phi_S$, if one knows the source of the driving force, we may further split velocities into $\nabla\phi_{u'} = \nabla\phi_u + \alpha_S\nabla\phi_S$, where $\alpha_S$ denotes the strength of the signal. Assume a cluster of food ($\rho_S(\mathbf{x})$) that releases any biosignatures such as heat, chemicals or biomolecules through Brownian diffusion at a slow enough rate such that the dimensionless concentration, $\phi_S$ follows $(\partial_{xt} + \nabla^2)\phi_S = \rho_S$. If the food is rigid and concentrated at a small enough region, $\nabla\phi_S$ becomes divergence free as $\partial_{xt} = 0$ and $\rho_S = 0$ outside the negligible region. Straightforwardly, all interaction in our framework update to include additional terms (see SI):

\begin{figure}[h]
\centering
\includegraphics[width=0.8\textwidth]{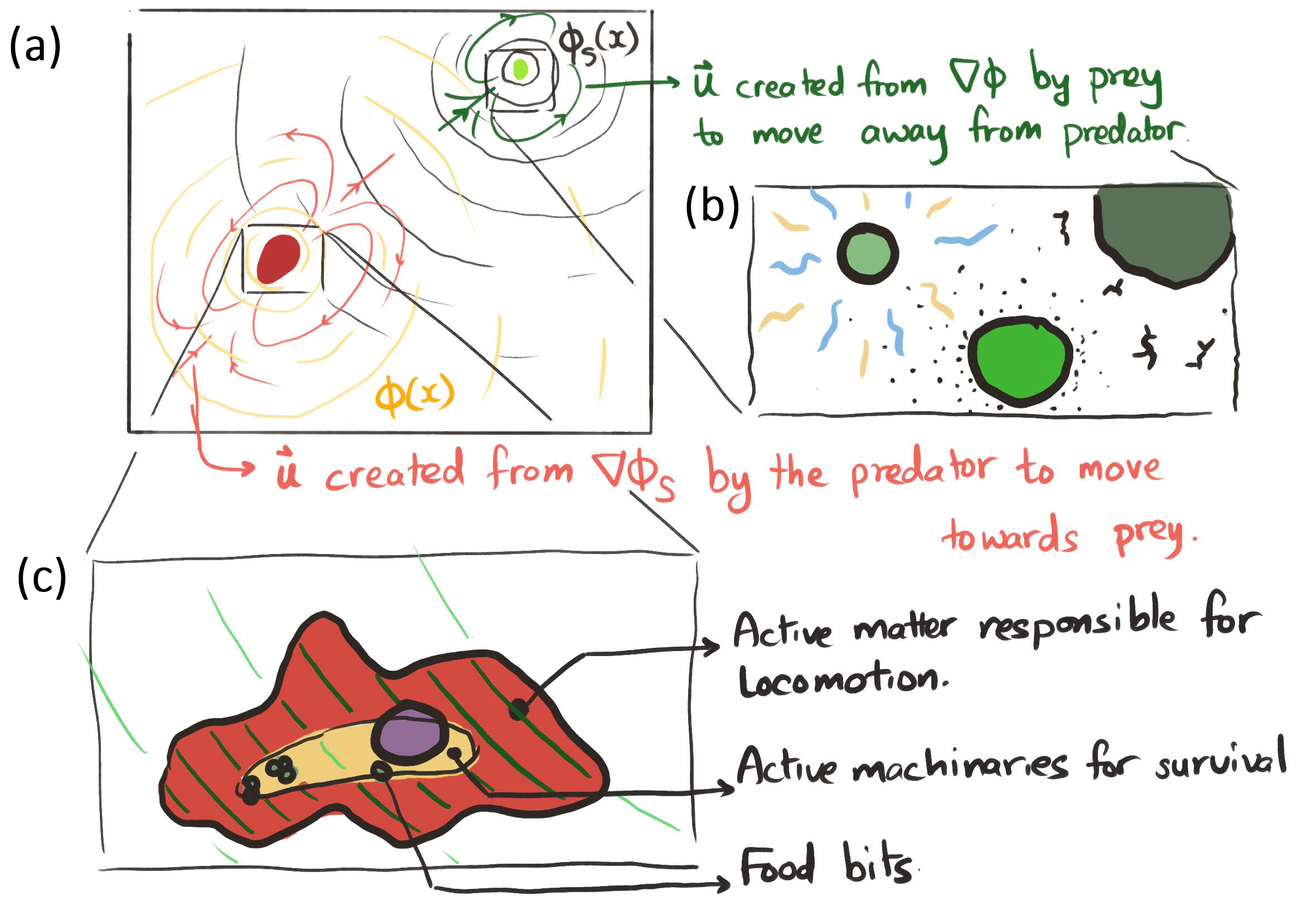}
\caption{Illustration of predator-prey operating on chemotaxis. (a) The biosignal fields $\phi_s$ and $\phi$ signaling predators and prey the best suited direction of motion to hunt/survive. (b) Typical exaple of biosignals, such as heat, sound, ATP and pheromones. (c) Internal arrangement of the predator that uses $\nabla \phi$ to generate activity (red), which also has activity not responsible for motion (orange).
}\label{fig3}
\end{figure}

\begin{equation}
\begin{cases}
    F_{u\phi}=\hbar_{A}\alpha_S\nabla\phi_S \cdot \nabla\phi \\
    f_{\phi_S\rho} = 0 \\
    \mathbf{f}_{\phi_S u} = -\rho_\psi q_S\nabla\phi_S \\
    f_{\phi_S Q} = \alpha_S\nabla \cdot (\mathbf{Q}\nabla\phi_S) + \widehat{\mathbf{S}}|_{\nabla\phi_S}(\mathbf{Q})
\end{cases}
\label{a2}
\end{equation}

In eq. (16), $\hbar_{A}\alpha_S\nabla\phi_S \cdot \nabla\phi$ is exactly the field energy stored between the predator and prey, $-\rho_\psi q_S\nabla\phi_S$ is the field force dragging our wavefunction ($\psi$ through strength $qm_\psi$) down the energy landscape, i.e., down the potential gradient $\phi_S$. Here the charge $q_S$ represents the affinity of the prey $\phi_S$ to the predator $\psi$, which is not necessarily equal to $q_\phi$. 

This is Gauss’s Law for Active Matter in disguise that changes the energy landscape for the survival of the active predator. Similarly, if the prey is modeled through the wavefunction, the affinity $q_{\phi S}$ becomes negative for the survival of prey and zero for passive food. Interestingly, if both predator-prey are modeled as two states of the same wavefunction, the symmetry of the Gauss’s law at quantum scale breaks. As the predator is chasing an escaping prey, the combined center of mass is moving forward. This mathematical result may explain why in a chase, predator and preys never move around any common point. Lastly, the $f_{\phi_S Q}$ is the term that tells the active machinery how to evolve in fulfillment of the task. Such fulfillment may be achieved through peristaltic motion as seen in the last section or any number of ways active creatures move. This is the point where the nematic ordering adds a crucial facet to AQN. Defects in nematic liquid crystals are of two types, the +1/2 comet and -1/2 star. The comet is a moving charge (predator) that has an affinity for the star (passive food), and as shown by Landau de Gennes at the dawn of Liquid Crystal theory, the attraction between the two follow the Gauss law as well \cite{shankar2018defect}. This facet of AQN points towards the Second Quantization, which is currently out of scope of this work.

\newpage
\subsection{Spherical Harmonics of Pulsating Bodies}\label{subsec3_3}

In the philosophy of quantum mechanics, the wavefunction is everything that could be meaningfully asked about the dynamics of a system. Learned we are that quantizing active matter is possible, let us look at approximately spherical pulsating active clumps. Such might be a fetal cell ball, a cancer growth, or a simplified heart that is active only over its surface. In fig. (xyz), we show a simplified heart which has a map of local ordering as $\vb{Q}_h$ varying negligibly in the material coordinate frame $\partial_{Xt}\vb{Q}_h$. The activity operator transforms our wavefunction according to $\hat{\zeta}_h\ket{\psi}=\ket{\zeta_h}$. The question now becomes, how would one find such eigenfunctions.

\begin{figure}[h]
\centering
\includegraphics[width=0.8\textwidth]{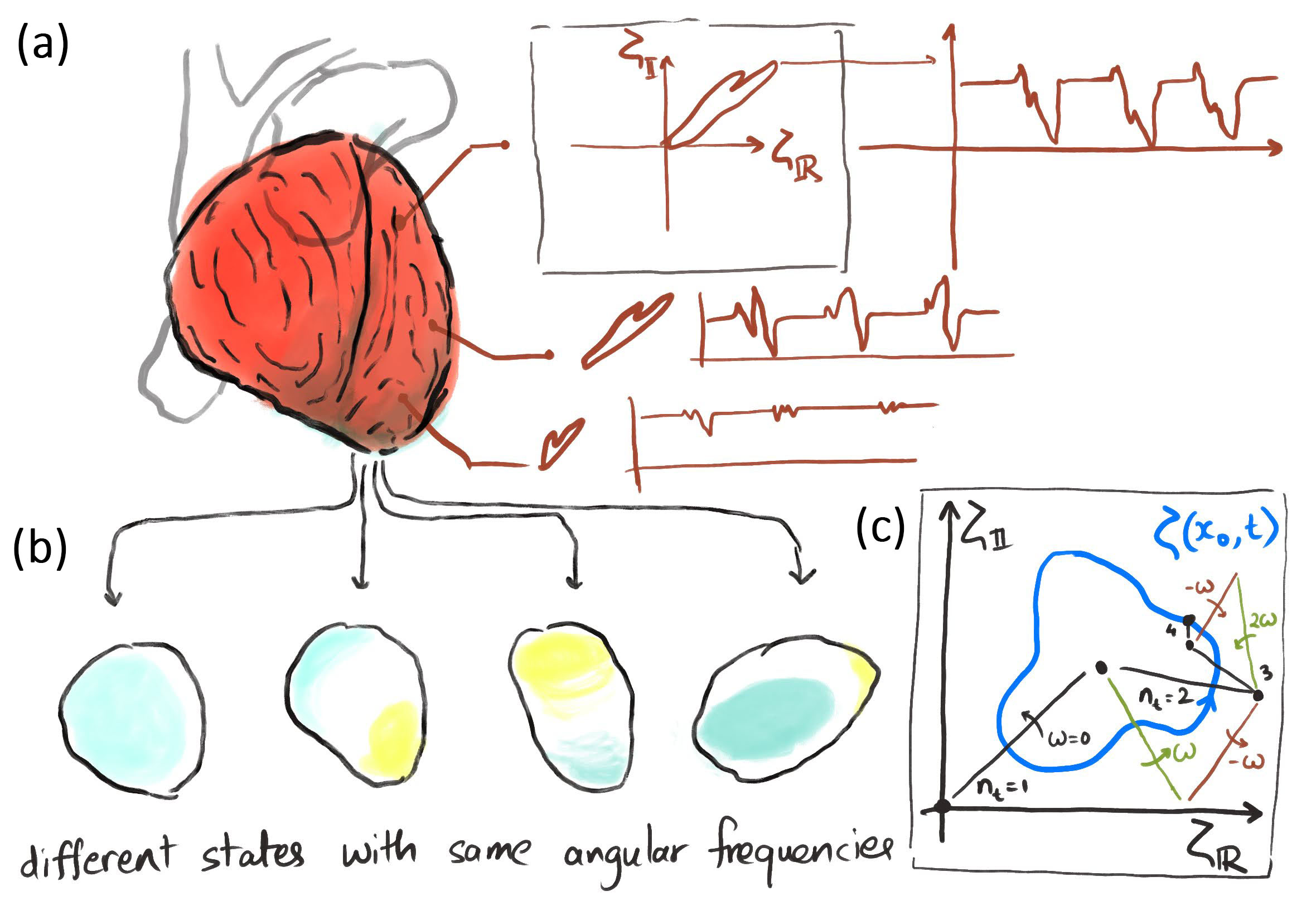}
\caption{ Illustration of pulsating active bodies. (a) A beating heart simplified to spherical bodies, with the pulses illustratively plotted in complex plane as well as pulse-time graph. (b) For any given energy eigenstates, various orbitals that beat independently to squeeze blood across body. (c) An illustration of how superposition of various eigenstates relates to evolving vectors in Hilbert space ($\ket{\zeta}$ in this particular figure).
}\label{fig4}
\end{figure}

Similar to the local phase variation of worm, let’s tracks the activity over time, and note the starting phase and the magnitude at various locations over the heart. Experimentally, this means recording electrical signals over the heart muscles at multiple locations. For some given beats per minute, we label our fundamental frequency $\omega_h=2\pi\times\text{BPM}/60$ and expand our system in its multiples as $n\omega_h$. We see different locations in the domain not only show a locally varying amplitude, but also a starting phase angle. Fourier transforming $\zeta(\vb{x},t)$ at each discrete harmonics over continuous space results in:

\newpage
\section{Closing Remarks}\label{conclusion}
Here we have shown how the use of the Dirac equation is not only exclusive to high energy, relativistic phenomenons, but also the most appropriate way to fully Fourier transform any active system. In cases where the amount of activity is not constant in time, where the amount of life is changing with time, the eigenvalues become imaginary and factors such as $\dot{\zeta}(\vb{x},t)$ are of importance. These would be terms that make life hostile or create resources and acting at longer spacio-temporal wavelengths. Renaming the quantized activity $\ket{\zeta}$ to 'life force' could be pedagogically compelling. 

Despite its simplicity, this framework is a shared foundation across many different fields. It unifies major fields of active matter physics such as elasticity of musculature, multi-phase boundaries, low Reynolds number fluid mechanics, active nematics and many other areas of elasto-porous hydrodynamics in cases where the dynamics of the fields are generated from life forces. Furthermore, pulsating, growing, 'dying', moving, sensing are all facets of every-day life around us that follow oscillatory and exponential dynamics. While applying tools from quantum mechanics, the exact distinction between polar, nematic, trigonal built of continuum is of less relevance as long as there is any information stored in the Fourier modes of its dynamics.

Lastly, Fourier transform allows us to truncate signals and effectively take a low, high or an arbitrary pass filters of the life force. While each pair of microtubules that make active nematics is incapable of generating a net force, we showed how peristaltic creature swim in fluids to hunt and survive. This is because we mathematically assigned the $\vb{Q}$ evolution as identically zero, and assumed them to be a mere muscle mechanism to used by high frequency signal generating machinery such as cilia hairs, flagella helicity, undulating motion of guts, fins of fish, etc. 

The same framework can be used to study in which quantum modes the  heart beats to pump blood with various degree of efficiency. There is no reason mathematically to base eigenstates $\omega_h$ in the multiple of BPMs. How heartbeats change over seconds of experiencing fear, minutes of running, a day of work and years of constant applications of stimuli are valuable quantities of interest. Here, the wavefunction does not define merely the fluid mechanics of the blood around an individual's circulatory system, but also allows us to store the sound and shape of the heart at inaudible frequencies, analyze, and transform the evolution of the life force back in the temporal space. We need not respect momentum conservation if we abstract out that the viscous dissipation are balanced by active energy. The energy in this framework refers to a scalar field that statistically models where the system wants to go and can take many meanings. At last, The life force is not a conserved property, it depends on the efficiency with which a creature or an organ utilizes resource around it towards their programed goals. Unlike the Aufbau principle, we expect occupation of various energy eigenstates to follow laws of evolution, i.e., achieving the goal of maximal resource collection with minimal energy expenditure.

\bibliography{sn-bibliography}% common bib file

\newpage
\begin{appendices}\label{appendix}

\section{}
\subsection{Deriving the Wave equation}\label{appendix1}
We start by assigning arbitrary patches of active regions with complex numbers as $\psi(\vb{x},t)$. These patches are expected to represent active flocks, creatures and organs. In fluid mechanics, such variables are transported through the transport equation (eq. \ref{eq:passive_patch}). 
\begin{equation}
    \partial_t \psi + \nabla \cdot (\vb{u} \psi) - D_\psi \nabla^2 \psi = S_\psi \label{eq:passive_patch}
\end{equation}

with diffusivity $ D_\psi $ and source term $ S_\psi $. Although species transport can easily be solved over complex domain, there is no meaningful interaction between the two complex components. Furthermore, we intend to conserve $\abs{\psi}^2$, which in general is contradicted by conserving a complex valued $\psi$. We do not intend our labels to diffuse over time, nor getting created or destroyed by any arbitrary
sources. Hence, we put $D_\psi=0$ and $S_\psi=0$. This gives us our reduced transport equation as $\partial_t \psi + \nabla \cdot (\vb{u} \psi) =0$ or $\partial_{Xt}\psi=0$ where any function of $\psi$ is now conserved because $\psi\p{\vb{X},t}=\text{const.}$. However, there is no meaningful interaction between the two complex components. With appropriate labels over our patch being transported as desired, we can start assigning meaning to the numerical values of the complex field $\psi$. Inspirationally from quantum mechanics, we assign $|\psi(\vb{x},t)|^2$ as the amount of presence at any point in space or time. 

If the complex phase could periodically change with time, we can assign motion of any oscillating creatures (such as the helical tail of sperm cell, the rhythmic ciliary motion, etc.) to the complex phase angle as a 'clock' for its creature.  A field with fixed magnitude in material frame but rotating in complex plane evolves as:
\begin{equation}
    \partial_{Xt} \psi = 0 \rightarrow \partial_{Xt}\psi = \text{i}\p{\omega_0+\omega}\psi \label{eq:rotating_phase}
\end{equation}

Here, $\omega(\vb{x},t)$ is the speed of the internal clocks of the creature at a given location at a given instant and $\omega_0$ is defined in the main text. Similarly, a static field that varies its phase over a length $\lambda_\phi=2\pi/|\vb{k}|$ (the de Broglie wavelength) in the direction $\hat{\vb{k}}$ changes its phase when moving in space with velocity $\vb{u}$ as $\partial_{Xt}\psi = \text{i}\vb{u}\cdot\vb{k}\psi$ and the combined equation becomes:

\begin{equation}
    \partial_{Xt}\psi = \text{i}\p{\omega_0+\omega}\psi \rightarrow \partial_{Xt}\psi = \text{i}\p{\omega_0 + \omega + \vb{u}\cdot\vb{k}}\psi
    \label{eq:translating_phase}
\end{equation}

Notice, eq. (\ref{eq:translating_phase}) is eq. (\ref{q1}) in the main text. By transformation of eq. (\ref{eq:passive_patch}) to eq. (\ref{eq:translating_phase}), $|\psi|^2$ is automatically conserved, as shown by ($\psi^* \times$(\ref{eq:translating_phase}) $+ \ \psi\times$(\ref{eq:translating_phase})$^*$), where $\p{}^*$ is the complex conjugation:
\begin{equation}
    \begin{array}{r@{\;}l}
        \psi^* \partial_{Xt} \psi &= \text{i}\p{\omega_0 + \omega + \vb{u}\cdot\vb{k}}|\psi|^2 \\
       \psi \partial_{Xt} \psi^* &= -\text{i}\p{\omega_0 + \omega + \vb{u}\cdot\vb{k}}|\psi|^2 \\
        \hline
        \partial_{Xt}(\psi^*\psi) &= 0
    \end{array}
    \label{conserve_psi}
\end{equation}

Which clearly shows $\partial_{Xt}(\psi^*\psi)=\partial_{Xt} (|\psi|^2) = 0$, hence, $|\psi(\vb{X},t)|^2 = \text{const.}$ over all characteristic curves, $\vb{X}(\vb{x}_0,t) = \vb{x}_0 + \int_t\vb{u}(\vb{X},t)\text{d} t$.
Here $\vb{X}(\vb{x}_0,t)$ is an equation of a streamline in the spatial coordinates such that $\vb{X}|_{t=0} = \vb{x}_0$. Ultimately, no matter what $\br{\omega+\vb{u}\cdot\vb{k}}(\vb{x},t)$ is, $|\psi|^2 = \rho_\psi$ and $\int_V \rho_\psi dV$ are conserved by construction (the new ``force" is simply rotating the complex phase).

\subsection{Deriving the Interaction Terms from Gauge Transformation}\label{appendix2}

Laws of physics must not change with our choice of reference frames. This principle gave rise to the relativistic and gauge theoretic description of physics in the last century. Imagine a creature swimming in a fluid bath. In the external reference frame, there is a force (could be external or self-propulsion) that moves the creature in space with time. However, in the reference frame of the creature, there is no propulsion. The creature can consistently believe that it is stationary in space  generating a self-flow that attracts (or repel) its food (or toxins) and the laws of physics must still hold. As discussed in the main paper, this is a simple application of Galilean relativity for active creatures.

We make a stronger statement here that the laws of physics (the governing equation, eq. (\ref{q1_2}-\ref{q4})) must remain valid for both active (moving) and inactive (resting) states of creatures. The problem at hand with this constraint is that manipulating $\psi$ through operators (such as $\hat{\dot{\rho}}$ or $\hat{\zeta}$) erases the information about the phase $\phi$. This is because all operators must be pre-multiplied with $\psi^*$ to bring the state vector $\ket{\psi}$ back into physical (real) space before being incorporated in our classical conservation laws. 

Nevertheless, we must address this conundrum as it is the complex phase that changes when the creature decides to become more or less active. In other words, the energy and momentum rate are non-zero ($E=\hbar_A\omega$ and $\vb{P}=\hbar_A\vb{k}$) for creatures in excited eigenstates, and non-zero $\omega,\vb{k}$ are bound to evolve $\phi$ in space and time. Conversely, if there is no energy-momentum consumption-generation, all interaction terms must be zero, i.e., $\cbr{\omega,\vb{k}}=0\, \Rightarrow\cbr{f_{\phi\rho}, \vb{f}_{\phi u},\vb{f}_{\phi Q}}=0$. This is straightforwardly true because if not for $f_{\phi\rho}, \vb{f}_{\phi u}$ and $\vb{f}_{\phi Q}$, eq. (\ref{q1_2}-\ref{q4}) are modeled after passive, porous, floating materials whose properties (such as $c_D, \vb{\Pi}^e,\zeta$, etc.) are stored in eigenvalues of various operators.

Next, we faithfully assume $\phi$ to be physical and demand our equations to be invariant under any choice of $\phi$. This is essentially the exercise of making our governing equations invariant under gauge transformations, namely, local $U(1)$ phase symmetry. We shall ask whether the exercise is meaningful or not later. If no coupling functions of the form $f(\psi^*\hat{f}\psi)$ are capable of introducing $\phi$ into the conservation of $\rho ,\vb{u}$ and $\vb{Q}$ through spatiotemporally varying real-valued distributions defined by any function $f$ and operator $\hat{f}$, then $\phi$ must already be a physically meaningful quantity in our macroscopic laws of nature.

Starting with a constant phase $\phi(\vb{x},t)=\phi_0$ (i.e., inactive creatures), we get $\psi_0={\psi_r(\vb{x},t)}\text{e}^{\text{i}\p{\omega_0t+\phi_0}}$ for an entirely real valued function $\psi_r$ and we call the background velocity $\vb{u}_0$. Substituting these fields in eq. (\ref{q1}) gives us:
\begin{equation}
    \begin{array}{cc}
         \quad & \textcolor{red}{\text{e}^{\text{i}\phi_0}}\partial_t \psi_r + \textcolor{blue}{\text{e}^{\text{i}\phi_0}\omega_0\psi_r} + 
     \textcolor{red}{\text{e}^{\text{i}\phi_0}}\nabla\cdot(\vb{u}_{0} \psi_r) = 
     \text{i}\p{\textcolor{blue}{\omega_0} + \omega + \vb{u}_0\cdot\vb{k}} \psi_r\textcolor{red}{\text{e}^{\text{i}\phi_0}},\\
     
         \quad & \partial_t \psi_r + 
     \nabla\cdot(\vb{u}_0 \psi_r) = 
     \text{i}\p{\omega + \vb{u}_0\cdot\vb{k}} \psi_r.
    \end{array}
    \label{charge_cons}
\end{equation}

In eq. (\ref{charge_cons}), LHS is purely real and RHS is purely imaginary and hence both must be individually zero. Consequently, $\partial_{Xt}\abs{\psi_0}=0$ from the real part. Furthermore, the imaginary part must be zero as well for all background velocities $\vb{u}_0$. Hence, $\omega=0,\vb{k}=0$ individually, which  matches our intuition about inactive creatures having BMR$=\hbar_A\omega$ and zero momentum-rate. The governing equations (eqs. (\ref{q2}-\ref{q4})) for passive flow read as:

\begin{equation}
\begin{array}{cc}
    & \textcolor{red}{\partial_{xt} \rho = \psi^* \hat{\dot{\rho}}\ \psi}, \\
    & \textcolor{red}{\partial_{xt}(\rho \vb{u}_0) = -\nabla p + \mu\nabla^2\vb{u}_0 + \mu_2 \nabla(\nabla\cdot\vb{u}_0) + \nabla\cdot (\mathbf{\Pi}^e + \psi^* \hat{\zeta}\, \psi \mathbf{Q})}, \\
    & \textcolor{red}{\partial_{xt} \mathbf{Q} = \Gamma \mathbf{H}}.
\end{array}
\end{equation}

Now, let's say the creature becomes active and builds a compass, $\phi=\phi(\vb{x},t)$ of the resource landscape of its surroundings. The new quantities are now $\psi=\psi_0\text{e}^{\text{i}\phi(\vb{x},t)}$, $\rho$, $\vb{u}$ and $\vb{Q}$, and the active velocity in eq. (\ref{q3}) is naturally obtained as $\vb{u}_A=\vb{u}-\vb{u}_0$. In the frame of active creatures, 

\begin{equation}
    \begin{array}{cc}
        & \p{\partial_{t} + \nabla\cdot\vb{u}} \psi_0\text{e}^{\text{i}\phi(\vb{x},t)}= \text{i}\p{\omega + \vb{u}\cdot\vb{k}}\psi_0\text{e}^{\text{i}\phi(\vb{x},t)}, \\
        
        & {\text{e}^{\text{i}\phi}\p{\partial_{t} + \nabla\cdot\vb{u}}\psi_0} + \psi_0 \p{\partial_{t} + \nabla\cdot\vb{u}} \text{e}^{\text{i}\phi} = \text{i}\p{\omega + \vb{u}\cdot\vb{k}}\psi_0\text{e}^{\text{i}\phi}, \\
        
        & \textcolor{red}{\text{e}^{\text{i}\phi}\partial_{Xt}\psi_0} + \textcolor{orange}{\psi_0\text{e}^{\text{i}\phi}} \p{\text{i}\partial_{t}\phi + \text{i}\vb{u}\cdot\nabla\phi} = \text{i}\p{\omega + \vb{u}\cdot\vb{k}}\textcolor{orange}{\psi_0\text{e}^{\text{i}\phi}}, \\
    \end{array}
    \label{phase_cons}
\end{equation}
where the first term is identically zero from eq. (\ref{charge_cons}). Since the remaining terms must be balanced for all velocities $\vb{u}$, the quantum properties must equate to phase angle as $\omega=\partial_t\phi$ and $\vb{k}=\nabla\phi$ as reported in eq. (\ref{q_int}). The phase is now rotating and repeating over time and space for non-zero energy-momentum-rate of creatures. If $\psi$ is used to describe physical objects (swimmers), the convection must emerge from a physical velocity by necessity. I.e., if the creature moves through swimming motion, the force from $\phi(x)$ should originate entirely from flow velocity. From Helmholtz decomposition, any velocity $\vb{u}_0 = \nabla\phi_u + \nabla\times\vb{\Psi}_u$. 

When a local phase variation is added, for the governing equations to stay valid, $\vb{u}_0 \rightarrow \vb{u}$ through a coupling constant $\alpha$ and the new velocity becomes $\vb{u} = \vb{u}_0 + \alpha\nabla\phi$ and thus, interaction terms arise when the creature adapts to move in environment of heterogeneous distribution of energy as dictated by $\phi(\vb{x},t) \neq \phi_0$. Substituting  $\vb{u} = \vb{u}_0 + \vb{u}_A=\nabla\phi_u + \nabla\times\vb{\Psi}_u +\alpha\nabla\phi$ in the governing equations gives:

\begin{equation*}
    \textcolor{red}{\partial_t \rho + \nabla \cdot (\vb{u}_0 \rho)} + \alpha \nabla^2 (\rho\phi) = \textcolor{red}{\psi^* \hat{m} \psi} + f_{\phi\rho}
\end{equation*}

\begin{equation*}
    \begin{gathered}
        \textcolor{red}{\partial_t(\rho\vb{u}_0)} + \alpha\partial_t(\rho\nabla\phi) + \textcolor{red}{\nabla \cdot (\vb{u}_0\vb{u}_0\rho)} + \textcolor{my_green}{2\alpha\nabla \cdot (\vb{u}_0\nabla\phi\rho)} + \textcolor{my_green}{\alpha^2\nabla \cdot (\nabla\phi\nabla\phi\rho)} \\
        = \\
        \textcolor{red}{-\nabla p} + \textcolor{red}{\mu\nabla^2\vb{u}_0} + \mu\nabla^2(\nabla\phi) + \mu_2\nabla(\nabla^2\phi) \\
        + \textcolor{red}{\nabla \cdot (\mathbf{\Pi}^e + \psi^*\hat{\zeta}\psi\mathbf{Q})} + \rho_\psi \alpha c_D\nabla\phi + \vb{f}_{\phi u}
    \end{gathered}
\end{equation*}

Therefore, $f_{\phi\rho}=\alpha \nabla^2 (\rho\phi)$. If the Reynolds numbers from both $\vb{u}$ and $\alpha\nabla\phi$ are low, the \textcolor{my_green}{non-linear} terms can be straightforwardly ignored, and $\vb{f}_{\phi u}$ becomes:
\begin{equation}
    \vb{f}_{\phi u} + \rho_\psi \alpha c_D\vb{u}_A = \alpha\nabla [\partial_t \rho - (\mu + \mu_2)\nabla^2]\phi + \rho_\psi \alpha c_D \nabla\phi.
\end{equation}

Similarly, $\mathbf{f}_{\phi Q} = \alpha \nabla \cdot (\mathbf{Q} \nabla \phi) + \hat{S} \big|_{\alpha\nabla\phi}(\mathbf{Q})$.

\subsection{Generalized Uncertainty Principles}\label{appendix3}

With the example of pressure gradient $\nabla p$, we demonstrate that the pressure force is simply $\mathbf{F}_p = -\langle\psi|\nabla p|\psi\rangle = -\int_V \rho_\psi\nabla p \,\mathrm{d}V$. Furthermore, if $\psi(\mathbf{x},t) = 1$ over any continuous patch of volume $V$, Stokes theorem returns the classical, $\mathbf{F}_p = -\oint_{\partial V} p\mathbf{n}_s \,\mathrm{d}x^2$, with an outward normal vector $\mathbf{n}_s$ to the surface $\partial V$. If two operator (let's say $\hat{A}, \hat{B}$) commute, $[\hat{A}, \hat{B}] = 0$, both operator share the same vector-space, but if the operators do not commute, the vectors $\hat{A}|\psi\rangle = |A\rangle$ and $\hat{B}|\psi\rangle = |B\rangle$ live in two different vector space and the generalized uncertainty principle holds:

\begin{equation}
\begin{cases}
    2\Delta A\Delta B \ge |\langle A|B\rangle| \\
    \Delta A = \sqrt{\langle\psi|\hat{A}^2|\psi\rangle - \langle\psi|\hat{A}|\psi\rangle^2} \\
    \Delta B = \sqrt{\langle\psi|\hat{B}^2|\psi\rangle - \langle\psi|\hat{B}|\psi\rangle^2}
\end{cases}
\label{guc}
\end{equation}

To understand eq.~(\ref{guc}), assume the growth rate of bacterial colony $M(\mathbf{x})$ and the activity $\zeta(\mathbf{x})$ of the corresponding continuum being linear with the concentration of ATP, implying $[\hat{M}, \hat{\zeta}] = 0$. Contrastingly, picture a multi-phase system with non-commuting mass density and viscosity for different non-linear relationships with phase fraction. Hence, in the distributions $2\Delta\rho\Delta\mu \ge |\langle\rho|\mu\rangle|$. Ultimately, the spectral theorem guarantees us a complete basis set $|A_n\rangle$ for any Hermitian operator $\hat{A}$. Therefore, $\psi(\mathbf{x})$ is a tool for modelling spatial variation of quantities, say $A(\mathbf{x}) = \Sigma a_n\hat{A}|\psi_n\rangle$, where $a_n$ are complex valued coefficients defining the state of our systems and $|a_n|^2$ gives the fraction of a given state over the total quantity in our object. Parallelly, $\omega$ allows us to capture the time variation of the system, which we demonstrate in the next section. Lastly, the interaction terms $\mathbf{f}_{\phi\rho}$, $\mathbf{f}_{\phi u}$ and $\mathbf{f}_{\phi Q}$ act through compositions of $\phi$ and $\rho, \mathbf{u}, \mathbf{Q}$ respectively and are defined in the next section.

 Let's take $\Delta t$ and $\Delta E$ as the standard deviation of the creature's lifespan and from the uncertainty relationship (see section \ref{appendix3}),
\begin{equation}
     2\Delta t \Delta E \geq \hbar_A
     \label{rol}
\end{equation}

This is exactly the hypothesis of the Rate-of-living theory.

\end{appendices}
\end{document}